\documentstyle{article}                     
\bibliography{plain}
\title{\bf Generalized Master Function Approach to Quasi-Exactly Solvable Models
}
\vspace{20mm}
\author{M. A. Jafarizadeh$^{a,b,c}$ \thanks{E-mail:jafarzadeh@ark.tabrizu.ac.ir} , S. J. Akhtarshenas  $^{a,b,c}$ \thanks{E-mail:akhtarash@ark.tabrizu.ac.ir}
\\
\\
$^a${\small Department of Theoretical Physics and Astrophysics, Tabriz University, Tabriz 51664, Iran.} \\   
$^b${\small Institute for Studies in Theoretical Physics and Mathematics, Tehran 19395-1795, Iran.} \\
$^c${\small Pure and Applied Science Research Center, Tabriz 51664, Iran.}}
\pagebreak

\begin{document}
\maketitle
\vspace{15mm}
\newpage
\begin{abstract}

By introducing the generalized master function of order up to four together
with corresponding weight function, we have obtained all quasi-exactly
solvable second order differential equations. It is shown that these
differntial equations have solutions of polynomial type with  factorziation
properties, that is polynomial solutions $P_m(E)$ can be factorized in terms
of polynomial $  P_{n+1}(E)$ for $m\geq n+1$. All known quasi-exactly quantum
solvable models can be obtained from these differential equations, where
roots of polynomial $  P_{n+1}(E)$ are corresponding eigen-values.


{\bf PACs Index: 03.65.Ge  }

\end{abstract}
\pagebreak

\vspace{7cm}

\section{INTRODUCTION}
During the last decade a remarkable new class of quasi-exactly solvable
spectral problems was introduced \cite{turb, shi-tur, ushv, shif, ul-yzal}.
These occupy an intermediate position
between exactly solvable and unsolvable models in the sense that exact solution in
an algebraized form exists only for a part of the spectrum.

The usual approach to the analysis of quasi-exactly solvable systems is an
algebraic one in which the operator is expressed as a non-linear combination
of generators of a Lie algebra. Another recent developement is the work of
Bender-Dunne \cite{ben-dun} where they have shown that the eigen-functions of a quasi-exactly
solvable schrodinger equation is the generating function for a set of orthogonal
polynomials ${P_m(E)}$ in energy variable. It was further shown that,
these polynomials satisfy the  three-term
recursion relation. Also, all polynomials beyond a critical polynomial
$P_{m}(E)$ factorize into the product of polynomial $P_{n+1}(E)$ and another
arbitrary polynomial.

In this paper we suggest a generalization of Bender-Dunne approach
to all possible one-dimentional quasi-exactly second order differential
equations.

For this purpose, the succesful master function approach of references
\cite{jafa, jaf}
to exactly solvable models, is generalized to a master function of
order up to four which gives all possible one-dimensional quasi-exactly
solvable models, where Bender-Dunne model \cite{ben-dun} and Heun differential equation
\cite{wang} are among them.

The paper is organized as follows: In section II we show that we can
generalize the usual quadratic master function to a master  function
of at most four order polynomials, then the most general quasi-exactly solvable
differential operators related to generalized master function of degree
$k=3$ and $k=4$ are given, respectively.

In section III, expanding their solutions in powers of $x$, we get 3-term
and 4-term recursion relations among their coefficeints, where
Bender-Dunne factorization follows through imposing the quasi-exactly
solvability conditions of section II. At the end of this section we
list all possible related quasi-exactly solvable differential equations
for $k=3$ and $k=4$ in Tables I and II, respectively.

Finally at section IV, we derive all possible  one-dimensional quasi-exactly
solvable quantum Hamiltonian from the differential operators of section III,
via prescription of references \cite{jafa, jaf}, where we have listed
them at the end of section III, except for those which can given in terms
of elliptic functions.
Paper ends with a brief conclusion.

\section{QUASI-EXACTLY SOLVABLE DIFFERENTIAL EQUATIONS ASSOCIATED WITH
GENERALIZED MASTER FUNCTIONS}

\setcounter{equation}{0}
By generalizing master function of order up to two \cite{jafa, jaf}
to polynomial of
order up to k, together with the non-negative weight function $W(x)$, defined
at interval (a,b) such that $\frac{1}{W(x)}\frac{d}{dx}\left(A(x)W(x)\right)$ to be
a polynomial of degree at most $(k-1)$, we can define the operator

\begin{equation}
L=\frac{1}{W(x)}\frac{d}{dx}\left(A(x)W(x)\frac{d}{dx}\right)+B(x),
\end{equation}
where $B(x)$ is a polynomial of order up to $(k-2)$. The interval $(a,b)$
is chosen so that, we have $A(a)W(a)=A(b)W(b)=0$.

It is straightforward to show that the above defined operator $L$ is a self adjoint
linear operator which at most maps a given polynomial of order $m$ to another polynomial
of order $(m+k-2)$. Now, by an appropriate choice of $B(x)$ and weight
function $W(x)$, the operator $L$ can have an invariant subspace of polynomials
of order up to $n$. Then by choosing the set of orthogonal polynomials
$\{\phi_0(x), \phi_1(x), \cdots, \phi_n(x)\}$ defined in the interval $(a,b)$
with respect to the weight function $W(x)$:

\begin{equation} 
\int_{a}^{b}\phi_m(x)\phi_n(x)W(x)dx=0, \quad\mbox{for}\quad\ m=n
\end{equation}
as the base, the matrix elements of the operator $L$ on this base will have
the following block diagonal form:

\begin{equation}
L_{ij}=0,\quad \quad if \quad
\{i\leq n \;\;\;and\;\;\; j\geq n+1 \}\;\;\;or\;\;\;
\{ i\geq n+1 \;\;\;and\;\;\; j\leq n \}.
\end{equation}

Since, according to the well known theorem of orthogonal polynomials, $\phi_n(x)$ is
orthogonal to any polynomial of order up to $n-1$, therefore, for matrix $L$
we get
\begin{equation}
L=\left[\begin{array}{cc}M&0\\\noalign{\medskip}0&N
\end{array}\right],
\end{equation}
where $M$ is an $(n+1)\times(n+1)$ matrix with matrix elements
\begin{equation}
M_{ij}=\int_{a}^{b}dxW(x)\phi_i(x)L(x)\phi_j(x),
\quad i,j=0,1,2,...,n ,
\end{equation}
and $N$ is an infinite matrix element defined as above with $i,j\geq n+1$.

The block diagonal form of the operator $L$ indicates that by diagonalizing the
$(n+1)\times(n+1)$ matrix M, we can find $(n+1)$ eigen-values of the operator $L$
together with the related eigen-functions as  linear functions of orthogonal
polynomials $\{\phi_0(x), \phi_1(x), \cdots, \phi_n(x)\}$ .

In order to determine the appropriate $B(x)$ and $W(x)$ for a given  generalized
master function $A(x)$, we Taylor expand those functions:
\begin{equation}
A(x)=\sum_{i=0}^{k}\frac{A^{(i)}(0)}{i!}x^{i},
\quad\quad \mbox{where}\quad A^{(i)}(0)=\frac{d^iA(x)}{dx^i}\mid_{x=0} 
\end{equation}
\begin{equation}
\frac{\left(A(x)W(x)\right)^\prime}{W(x)}=
\sum_{i=0}^{k-1}\frac{\left(\frac{(AW)^\prime}{W}\right)^{(i)}(0)}{i!}x^{i},
\quad\quad\mbox{ where}\quad \left(\frac{(AW)^\prime}{W}\right)^{(i)}(0)
=\frac{d^i\left(\frac{(A(x)W(x))^\prime}{W(x)}\right)}{dx^i}\mid_{x=0} 
\end{equation}
\begin{equation}
B(x)=\sum_{i=0}^{k-2}\frac{B^{(i)}(0)}{i!}x^{i},
\quad\quad \mbox{where}\quad B^{(i)}(0)=\frac{d^iB(x)}{dx^i}\mid_{x=0}. 
\end{equation}
Then, the existence of invariant subspace of the polynomials of order $n$ of the
operator $L$ leads to the following linear equation between the coefficients
of above Taylor expansions:
\begin{eqnarray}
-\frac{A^{(i+2)}}{(i+2)!}\;l(l-1)
-\frac{{\left(\frac{(AW)^\prime}{W}\right)}^{(i+1)}}{(i+1)!}\;l
+\frac{B^{(i)}}{i!}=0,\\
\nonumber
\end{eqnarray}
where
\begin{eqnarray}
\nonumber
\left\{\begin{array}{cccccc}
l=n, & and & i=1, & 2, & ...,  & k-2\\
l=n-1, & and & i=2, & 3, & ..., & k-2\\
... & ... &... & ... & ... & ...\\
l=n-k+4, & and & i=k-3, & k-2  & & \\
l=n-k+3, & and & i=k-2 & & & \\
\end{array}
\right..
\end{eqnarray}

The number of above equations, for a given value of $k$, is $\frac{(k-1)(k-2)}{2}$.
If we are to determine only the unknown function $B(x)$ without having any further
constraint on the weight function $W(x)$, then the above $\frac{(k-1)(k-2)}{2}$
equations should be satisfied with $(k-2)$ coefficients of Taylor
expansion of $B(x)$ as the only unknowns, since $B^{(0)}$ can be absorbed in
the eigen-spectrum operator $L$. Therefore, we are left with $(k-2)$ unkowns to be
determined, where the compatibility of equations (2-9) require $k=3$ at most.
On the other hand, if we add the coefficients of Taylor expansions of $A(x)$
and $\frac{\left(A(x)W(x)\right)^{\prime}}{W(x)}$ to our list of unknowns,
( to be determined by solving equations (2-9) ),
then their compatibility conditions require that:
\begin{equation}
3(k-1)\geq\frac{(k-1)(k-2)}{2},
\end{equation}
or $k\leq 8$, where further investigations show that we can
have at most $k=4$, since for $k\geq5$ the coefficients $A^{(k)}(0)$ and
$\left(\frac{(AW)^{\prime}}{W}\right)^{(k-1)}(0)$ will vanish.
Below we summarize the above-mentioned discussion for $k=3$ and $k=4$,
separately.

\subsection{$k=3$}

In this case, $B(x)$ is a second order polynomial where $B^{(1)}$ can be
determined by solving equation (2-9): 
\begin{equation}
B^{(1)}=\frac{n}{2}\left( \frac{A^{(3)}}{3}(n-1)
+\left(\frac{(AW)^\prime}{W}\right)^{(2)}\right),
\end{equation}
which is the only unknown in this case.

\subsection{$k=4$}

Again, the solving of equations (2-9) leads to:
\begin{equation}
B^{(1)}=\frac{n}{2}\left( \frac{A^{(3)}}{3}(n-1)
+\left(\frac{(AW)^\prime}{W}\right)^{(2)}\right),
\end{equation}
\begin{equation}
B^{(2)}=-\frac{A^{(4)}}{12}n(n-1),
\end{equation}
and
\begin{equation}
\left(\frac{(AW)^\prime}{W}\right)^{(3)}=-\frac{A^{(4)}}{2}(n-1).
\end{equation}
Here, besides having constraint over second order polynomial $B(x)$,
we have to put further constraints on the weight function $W(x)$
given in (2-14).

Definetly we can determine $n+1$ eigen-spectrum of the operator $L$, simply by
diagonalizing the $(n+1)\times(n+1)$ matrix $M$, since it is a self-adjoint
operator in Hilbert space of polynomials and it has a
block diagonal form given in (2-4).

As we are going to see in the next section, we can determine its eigen-spectrum
analytically, using some recursion relations.

\section{RECURSION RELATIONS}

In this section we show that the eigen-functions of the operator $L$
are a generating function for a new set of polynomials ${P_m(E)}$
where the eigen-function equation of the operator $L$ leads to the
recursion relation between these polynomials. Quasi-exact solvable
constraints (2-9) will lead to their factorization,  that is, $P_{n+N+1}(E)=P_{n+1}(E)Q_{N}(E)$ for $N\geq0$,
where roots of polynomials $P_{n+1}(E)$ turn out to be the eigen-values
of the operator $L$.

To achieve these results, first we expand $\psi(x)$, the eigen-function of
$L$, as:
\begin{equation}
\psi(x)=\sum_{m=0}^{\infty}P_m(E)x^m,
\end{equation}
where eigen-function equation:
\begin{equation}
L\psi(x)=E\psi(x)
\end{equation}
can be expressed as:
$$
-A(x)\sum_{m=2}^{\infty}m(m-1)P_m(E)x^{m-2}
-\frac{{\left(A(x)W(x)\right)}^\prime}{W(x)}\sum_{m=1}^{\infty}mP_m(E)x^{m-1}
$$
\vspace{-5mm}
\hspace{20mm}
\begin{equation}
+B(x)\sum_{m=0}^{\infty}P_m(E)x^m
=E\sum_{m=0}^{\infty}P_m(E)x^m,
\end{equation}
which leads to the following recursion relations for the
coefficients $P_{m}(E)$:
$$
\left(A^{(1)}(m+1)(m+2)
+{\left(\frac{{(AW)}^\prime}{W}\right)}^{(0)}(m+2)\right)P_{m+2}(E)
$$
$$
+\left(\frac{A^{(2)}}{2!}m(m+1)
+{\left(\frac{{(AW)}^\prime}{W}\right)}^{(1)}(m+1)+E\right)P_{m+1}(E)
$$
$$
+\left(\frac{A^{(3)}}{3!}m(m-1)
+\frac{{\left(\frac{{(AW)}^\prime}{W}\right)}^{(2)}}{2!}m
-B^{(1)}\right)P_m(E)
$$
\vspace{-1mm}
\hspace{20mm}
\begin{equation}
+\left(\frac{A^{(4)}}{4!}(m-1)(m-2)
+\frac{{\left(\frac{{(AW)}^\prime}{W}\right)}^{(3)}}{3!}m
-\frac{B^{(2)}}{2!}\right)P_{m-1}(E)=0.
\end{equation}

Below we investigate recursion relations thus obtained for $k=3$ and $k=4$,
separately.

\subsection{$k=3$}

In this case the 4-term general recursion relation reduse to the following
3-term recursion relation:
$$
\left(A^{(1)}(m+1)(m+2)
+{\left(\frac{{(AW)}^\prime}{W}\right)}^{(0)}(m+2)\right)P_{m+2}(E)
$$
$$
+\left(\frac{A^{(2)}}{2!}m(m+1)
+{\left(\frac{{(AW)}^\prime}{W}\right)}^{(1)}(m+1)+E\right)P_{m+1}(E)
$$
\vspace{-1mm}
\begin{equation} 
+\left(\frac{A^{(3)}}{3!}m(m-1)
+\frac{{\left(\frac{{(AW)}^\prime}{W}\right)}^{(2)}}{2!}m
-B^{(1)}\right)P_m(E)=0.
\end{equation}

In order to have finite eigen-spectrum, that is, quasi-integrable differential
equation, the above recursion relation should be truncated for some value of
$m=n$, which is obviously possible by an appropriate choice of:
\begin{equation}
B^{(1)}=\frac{n}{2}\left( \frac{A^{(3)}}{3}(n-1)
+\left(\frac{(AW)^\prime}{W}\right)^{(2)}\right),
\end{equation}
which is in agreement with the result of previous section given in (2-11).

Using the recursion relation (3-19), with $B^{(1)}$ given in (3-20),
we get a factorization of polynomial
$P_{n+N+1}(E)$ for $N\geq0$ in terms of $P_{n+1}(E)$
as follows:
\begin{equation}
P_{n+N+1}(E)=P_{n+1}(E)Q_{N}(E)\;\;\;N\geq0
\end{equation}
where, by choosing the eigen-values $E$ as roots of polynomial $P_{n+1}(E)$,
all polynomials of order higher than $n$ will vanish.

In order to determine corresponding
eigen-functions, it is sufficient to evaluate $P_{m}(E_{i})$ for
$m=0,\;1,\;2\;...\;n$ with
$E_i$ as roots of $P_{n+1}(E)$, then eigen-function $\psi_i(x)$ corresponding
to eigen-value $E_i$ can be written as:
\begin{equation}
\psi_i(x)=\sum_{m=0}^{n}P_m(E_i)x^m, \quad\quad  i=0\;,1\;...\;,n.
\end{equation}

The above eigen-functions are polynomials of order $n$, hence they can have
at most $n$ roots in the interval $(a,b)$, where, according to the well known
oscillation and comparison theorem of second-order linear differential
equation \cite{codd-levi}, these numbers order the eigen-values according to the number of roots
of corresponding eigen-functions.  Therefore, we can say that the
eigen-values thus obtained are the first $n+1$ eigen-values of the operator $L$.

Using the recursion relations (3-19), we can evaluate the polynomials $P_m(E)$
in terms of $P_0(E)$, where we have chosen $P_0(E)=1$. We have evaluated
the first five polynomials which appear in the Appendix I.

As an illustration we give the results for
$A(x)=x$ and $n=3$ with $\alpha=1$, $\beta=0$, $\gamma=-1$ which is
equivalent to the Bender-Dunne model:
$$
P_1(E)=-\frac{1}{2}E,
$$\vspace{-2mm}
$$
P_2(E)=-1+\frac{1}{12}E^2,
$$\vspace{-2mm}
$$
P_3(E)=\frac{1}{4}E-\frac{1}{144}E^3,
$$\vspace{-2mm}
$$
P_4(E)=\frac{1}{10}-\frac{1}{48}E^2+\frac{1}{2880}E^4.
$$

Obviously $P_m(E)$ have the parity of $m$.

By finding the 4-roots of $P_4(E)$ we determine the corresponding four
eigen-values:
$$
E_0=-7.398556194,
$$\vspace{-6mm}
$$
E_1=-2.293766823,
$$\vspace{-6mm}
$$
E_2=2.293766823,
$$\vspace{-6mm}
$$
E_3=7.398556194.
$$\vspace{-6mm}

Finally for the coefficient $P_m(E_i)$ we get:

$$
P_0(E_0)=1,
\;\;P_1(E_0)=3.699278097,
\;\;P_2(E_0)=3.561552813,
\;\;P_3(E_0)=-.962769686,
$$
$$
P_0(E_1)=1,
\;\;P_1(E_1)=1.146883412,
\;\;P_2(E_1)=-.5615528135,
\;\;P_3(E_1)=-.4896337383,
$$
$$
P_0(E_2)=1,
\;\;P_1(E_2)=-1.146883412,
\;\;P_2(E_2)=-.5615528135,
\;\;P_3(E_2)=.4896337383,
$$
$$
P_0(E_3)=1,
\;\;P_1(E_3)=-3.699278097,
\;\;P_2(E_3)=3.561552813,
\;\;P_3(E_3)=-.962769686.
$$

Using the above coefficients we can determine the corresponding eigen-functions
through formula (3-22).

In Table I we give all quasi-exactly solvable operators which can be obtained
by choosing different generalized master function of order 3. This Table
contains all possible models corresponding to different choice of $A(x)$
up to translation and rescaling of variable $x$. Also by choosing $A(x)$
as a polynomial of up to second order with $\gamma=0$ we lead to the
exactly solvable models of references \cite{jafa, jaf}.

\begin{table}
\begin{tabular}{|c|c|c|} \hline    
 $A(x)$  &   $W(x)$ & $L(x)$ \\ \cline{2-2}
 &  $intervals$ &  \\  \hline
 $x$  &  $x^\alpha e^{\beta x+\gamma x^2}$
& $-x\frac{d^2}{dx^2}-(\alpha+1+\beta x$
\\  &  $0 \leq x<+\infty$ 
&$+2\gamma x^2)\frac{d}{dx}+2n\gamma x$
\\  & $\alpha>-1,\; -\infty<\beta<+\infty,\; \gamma<0$ & 
\\  \hline
 $x^2$  &  $x^\alpha e^{\beta/x+\gamma x}$
& $-x^2\frac{d^2}{dx^2}+(\beta-(\alpha+2)x$
\\  &  $0 \leq x<+\infty$ 
&$-\gamma x^2)\frac{d}{dx}+n\gamma x$
\\  & $-\infty<\alpha<+\infty,\; \beta<0,\; \gamma<0$ & 
\\  \hline
 $x(1-x)$ &  $x^\alpha(1-x)^\beta e^{-\gamma x}$ &
 $x(x-1)\frac{d^2}{dx^2}+(-\alpha-1$
\\  &  $0\leq x\leq+1$ & $+(\alpha+\beta+\gamma+2)x$
\\  & $\alpha>-1,\; \beta>-1,\; -\infty<\gamma<+\infty$ & $-\gamma x^2)\frac{d}{dx}+n\gamma x$
\\  \hline
$x^3$  &  $x^\alpha e^{-\beta/x^2-\gamma/x}$
&  $-x^3\frac{d^2}{dx^2}-(2\beta+\gamma x$
\\  &  $0\leq x<+\infty$ 
&$+(\alpha+3)x^2)\frac{d}{dx}+n(n+\alpha+2)x$
\\  & $\alpha<-3,\; \beta>0,\; -\infty<\gamma<+\infty$ & 
\\  \hline
$x^2(1-x)$ &  $x^\alpha(1-x)^\beta e^{-\gamma/x}$
&  $x^2(x-1)\frac{d^2}{dx^2}+(-\gamma+(\gamma-\alpha-2)x$
\\  &  $0\leq x\leq1$ 
&$+(\alpha+\beta+3)x^2)\frac{d}{dx}$
\\  & $-\infty<\alpha<+\infty,\; \beta>-1,\; \gamma>0$ & $-n(n+\alpha+\beta+2)x$
\\  \hline
$x(1+x^2)$ &  $x^\alpha(1+x^2)^\beta e^{\gamma tan^{-1}x}$
&  $-x(1+x^2)\frac{d^2}{dx^2}-(\alpha+1+\gamma x$
\\  &  $0\leq x<+\infty$ &
$+(\alpha+2\beta+3)x^2)\frac{d}{dx}$
\\  & $\alpha>-1,\; \beta<-(\alpha+3)/2,\; -\infty<\gamma<+\infty$ & $+n(n+\alpha+2\beta+2)x$
\\  \hline
$x(1-x)(a-x)$ &  $x^\alpha(1-x)^\beta(a-x)^\gamma$
&  $x(x-1)(a-x)\frac{d^2}{dx^2}+(-a(\alpha+1)$
\\ $a>1$  &  $0\leq x\leq+1$ &
$+((a+1)(\alpha+2)+a\beta+\gamma)x-(\alpha+\beta$
\\  & $\alpha>-1,\; \beta>-1,\; -\infty<\gamma<+\infty$ & 
$+\gamma+3)x^2)\frac{d}{dx}+n(n+\alpha+\beta+\gamma+2)x$
\\  \hline
\end{tabular}
\caption{Quasi-exactly differential operators obtained from generalized master
function of order up to 3 }
\end{table}

\subsection{$k=4$}

Again in order to truncate the recursion relation (3-18) and to factorize
polynomials $P_{n+N+1}(E)$ in terms of $P_{n+1}(E)$, we should have:

\begin{equation}
B^{(1)}=\frac{n}{2}\left( \frac{A^{(3)}}{3}(n-1)
+\left(\frac{(AW)^\prime}{W}\right)^{(2)}\right),
\end{equation}
\begin{equation}
\frac{B^{(2)}}{2!}=\frac{A^{(4)}}{4!}(n-1)(n-2)
+\frac{{\left(\frac{{(AW)}^\prime}{W}\right)}^{(3)}}{3!}n,
\end{equation}
and
\begin{equation}
\frac{B^{(2)}}{2!}=\frac{A^{(4)}}{4!}n(n-1)
+\frac{{\left(\frac{{(AW)}^\prime}{W}\right)}^{(3)}}{3!}(n+1).
\end{equation}

Solving the above equations we get:
\begin{equation}
B^{(2)}=-\frac{A^{(4)}}{12}n(n-1),
\end{equation}
and
\begin{equation}
\left(\frac{(AW)^\prime}{W}\right)^{(3)}=-\frac{A^{(4)}}{2}(n-1).
\end{equation}

The equations (3-23) , (3-26) and (3-27) are the same equations which
are required in the reduction of the operator $L$ to its block diagonal form.

Again roots of polynomials $P_{n+1}(E)$ will correspond to $n+1$ eigen-values
of the differential operator $L$ with eigen-functions which can be expressed in
terms of $P_{m}(E_i)$ for $m\leq n$, where polynomials $P_m(E)$ can be obtained
from recursion relation by choosing $P_0(E)=1$ and $P_{-1}(E)=0$, where
we have given the first four polynomials in Appendix II.

In Table II we list all quasi-exactly differential operators which can be
obtained from the generalized master function of order up to four.

\begin{table}
\begin{tabular}{|c|c|c|}     \hline
 $A(x)$  &   $W(x)$ & $L(x)$ \\ \cline{2-2}
 &  $intervals$ &  \\  \hline
 $x^4$  &  $x^\alpha e^{\beta/x^3+\gamma/x^2+\delta/x}$
& $-x^4\frac{d^2}{dx^2}+(3\beta+2\gamma x$
\\  &  $0 \leq x<+\infty$ 
&$+\delta x^2-(\alpha+4)x^3)\frac{d}{dx}-n\delta x-n(n-1)x^2$
\\  & $\beta<0,\; -\infty<\gamma<+\infty,\; -\infty<\delta<+\infty$ &
\\  & $\alpha=-2(n+1)$ & 
\\  \hline

 $x^3(1-x)$  &  $x^\alpha (1-x)^\beta e^{-\gamma/x-\delta/x^2}$
& $x^3(x-1)\frac{d^2}{dx^2}+(-2\delta+(2\delta-\gamma)x$
\\  &  $0 \leq x\leq +1$ 
&$+(\gamma-\alpha-3)x^2+(\alpha+\beta+4)x^3)\frac{d}{dx}$
\\  & $\beta>-1,\; -\infty<\gamma<+\infty,\; \delta>0$ & 
$+n(n+\alpha-\gamma+2)x+n(n-1)x^2$
\\  & $\alpha=-2(n+1)-\beta$  & 
\\  \hline

 $x^2(1+x^2)$  &  $x^\alpha (1+x^2)^\beta e^{\gamma/x+\delta tan^{-1}(x)}$
& $-x^2(1+x^2)\frac{d^2}{dx^2}+(\gamma-(\alpha+2)x$
\\  &  $0 \leq x<+\infty$ 
&$+(\gamma-\delta)x^2-(\alpha+2\beta+4)x^3)\frac{d}{dx}$
\\  & $-\infty<\beta<+\infty,\; \gamma<0,\; -\infty<\delta<+\infty$ & 
$+n(\delta-\gamma)x-n(n-1)x^2$
\\  & $ \alpha=-2(n+\beta+1)$  & 
\\  \hline

 $x^2(1-x)(a-x)$  &  $x^\alpha(1-x)^\beta(a-x)^\gamma e^{\delta/x}$
& $x^2(x-1)(a-x)\frac{d^2}{dx^2}+(a\delta-(a\alpha+(a+1)\delta$
\\ $a>1$ &  $0\leq x\leq+1$ 
&$+2a)x-((a+1)\alpha+a\beta+\gamma+\delta$
\\  & $\beta>-1,\; -\infty<\gamma<+\infty,\; \delta<0$ & 
$+3(a+1))x^2-(\alpha+\beta+\gamma+4)x^3)\frac{d}{dx}+n((\alpha$
\\  & $\alpha=-2(n+1)-\beta-\gamma$  &$-n+4)(a+1)+a \beta+\gamma+\delta)x-n(n-1)x^2$ 
\\  \hline

 $x^2(1-x)^2$  &  $x^\alpha(1-x)^\beta e^{\gamma/x+\delta/(1-x)}$
& $-x^2(1-x)^2\frac{d^2}{dx^2}+(\gamma-(\alpha+2\gamma+2)x$
\\  &  $0\leq x\leq +1$ 
&$+(2\alpha+\beta+\gamma-\delta+6)x^2-(\alpha+\beta+4)x^3)\frac{d}{dx}$
\\  & $-\infty<\beta<+\infty,\; \gamma<0,\; \delta<0$ & 
$-n(+2n+2\alpha+\beta+\gamma-\delta+4)x-n(n-1)x^2$
\\  & $\alpha=-2(n+1)-\beta$  & 
\\  \hline

 $x(a-x)(1+x^2)$  &  $x^\alpha(a-x)^\beta(1+x^2)^\gamma e^{\delta tan^{-1}x}$
& $x(x-a)(1+x^2)\frac{d^2}{dx^2}+(-a(\alpha+1)$
\\ $a>0$ &  $0\leq x\leq a  $ &
$+(\alpha+\beta-a(\delta-2))x-(a(\alpha+2\gamma+3)$
\\  & $-\infty<\gamma<+\infty,\; -\infty<\delta<+\infty$ &
$-\delta)x^2+(\alpha+\beta+2\gamma+4)x^3)\frac{d}{dx}$
\\ & $-1<\beta<-2n-2\gamma-1$  & $+n(a(n+\alpha+2\gamma+2)-\delta)x+n(n-1)x^2$
\\  & $\alpha=-2(n+1)-\beta-2\gamma$ & 
\\  \hline

$x(1-x)(a-x)(b-x)$ & $x^{\alpha}(1-x)^{\beta}(a-x)^{\gamma}(b-x)^{\delta}$
& $x(x-1)(a-x)(b-x)\frac{d^2}{dx^2}$
\\ $1<a<b$ &  $0\leq x\leq +1$ &
$+(-ab(\alpha+1)+(2(a+b+ab)+ab(\alpha+\beta)$
\\  & $-1<\beta<2n-\gamma-1$ &
$+b(\alpha+\gamma)+a(\alpha+\delta))x-((a+b)(\alpha+\beta+3)$
\\ & $\-\infty<\gamma<+\infty,\; -\infty<\delta<+\infty$ &
$+(a+1)\delta+\alpha+\gamma+3)x^2+(\alpha+\beta+\gamma+4)$
\\ & $\alpha=-2(n+1)-\beta-\gamma$  &$x^3)\frac{d}{dx}+n((a+b)(\alpha+\beta+n+2)$
\\  &  & $+(a+1)\delta+\alpha+\gamma+n+2)x+n(n-1)x^2$
\\  \hline
\end{tabular}
\caption{Quasi-exactly differential operators obtained from generalized master
function of order up to 4 }
\end{table}

Tables I and II contain all quasi-exactly second order differential
equations which can be obtained from Lie algebraic methods \cite{shif}.
For example we get the Bender-Dunne model \cite{ben-dun}
for the choice of $A(x)=x$ and $\beta=0$, and similarly,
we get the Heun differential operator ( Fuxian equation with four regular
singular point ) \cite{wang},
for the choice of $A(x)=x(x-1)(x-a)$.

\section{QUASI-EXACTLY POTENTIAL ASSOCIATED WITH GENERALIZED MASTER FUNCTION}

As in references \cite{jafa, jaf}, writting:
\begin{equation}
\psi(t)=A^{1/4}(x)W^{1/2}(x)\phi(x),
\end{equation}
with a change of variable $\frac{dx}{dt}=\sqrt{A(x)}$, the eigen-value
equation of the operator $L$ reduces to the schrodinger equation:
\begin{equation}
H(t)\psi(t)=E\psi(t),
\end{equation}
with the same eigen-value E and $\psi(t)$ given in (4-29), in terms of
the eigen function of $L$, where $H(t)=-\frac{d^2}{dt^2}+V(t)$
is the similarity transformation of $L(x)$ defined as:
\begin{equation}
H(t)=A^{1/4}(x)W^{1/2}(x)L(x)A^{-1/4}(x)W^{-1/2}(x)
\end{equation}
with:
\begin{equation}
V(t)=-\frac{3}{16}\frac{\dot{A}^2(t)}{A^2(t)}
-\frac{1}{4}\frac{\dot{W}^2(t)}{W^2(t)}
+\frac{1}{4}\frac{\dot{A}(t)\dot{W}(t)}{A(t)W(t)}
+\frac{1}{4}\frac{\ddot{A}(t)}{A(t)}
+\frac{1}{2}\frac{\ddot{W}(t)}{W(t)}+B(t).
\end{equation}

It is also straightforward to show that:
\begin{equation}
\int dt\phi(t)H(t)\psi(t)
=\int_{a}^{b}dxW(x)\psi(x)L(x)\psi(x).
\end{equation}

Hence block diagonalization of $L$ leads to block-diagonalization of $H$.

As an illustration we give below an example with $A(x)=x^3$, weight function
$W(x)=x^{\alpha}e^{\beta/x^2-\gamma/x}$ and interval $[0,\infty)$, where
$\alpha<-3$ and $\beta,\;\gamma >0$.

From a change of variable $dx/dt=\sqrt{A(x)}$, we get $x(t)=4/t^2$, hence using
equation (4-31) we have for potential $V(t)$:

$$
V_n(t)=\frac{\gamma}{2}\left(\alpha+1 \right )
+\left({\frac {15}{4}}+{\alpha}^{2}+4\,n\alpha+4\,\alpha+4\,{n}^{2}+8\,n \right)
\frac{1}{t^2}
+\frac{1}{4}\left (\alpha\,\beta+\frac{1}{4}\,{\gamma}^{2}\right ){t}^{2}
$$
\begin{equation}
+\frac{\beta\,\gamma}{16}{t}^{4}
+\frac {{\beta}^2}{64}{t}^{6}
\end{equation}

Below we give a list of quasi-exact solvable potentials, except for those
potentials which can be expressed in terms of elliptic functions, since
in this case we get rather long expressions for them:

$$     
A(x)=x, \quad x(t)=\frac{t^2}{4}
$$\vspace{-5mm}
$$
V_n(t)=\frac{\beta}{2}\left(\alpha+1 \right )
+\left(\alpha^2-\frac{1}{4} \right )\frac{1}{t^2}
+\frac{1}{2} \left (\frac{\beta^2}{8}
+\gamma \left( n+1+\frac{\alpha}{2} \right ) \right){t}^{2}
+\frac{\beta \gamma}{16}\;t^4
+\frac {\gamma^2}{64}\;t^6,
$$

$$
A(x)=x^2, \quad x(t)=e^t
$$\vspace{-5mm}
$$
V_n(t)=\frac{1}{4}\left (1+\alpha^2-2\beta \gamma+2\alpha
-2\alpha\beta e^{-t}
+{\beta}^{2}{e^{-2t}}
+{{\it \gamma}}^{2}{e^{2t}}
+2\left(2\;\gamma+2n \gamma+\alpha\;\gamma \right )e^t \right),
$$

$$     
A(x)=x(1-x), \quad x(t)=\frac{1+sin(t)}{2}
$$\vspace{-4mm}
$$
V_n(t)=\frac{1}{2}\left (+n\;\gamma-\alpha\,\beta-\beta-\alpha
+\frac{1}{2}\left(\beta\,\gamma-\alpha^2-\beta^2-\alpha\,\gamma-1 \right)
+\left (\frac{\alpha\,\gamma}{2}\,+\gamma+\frac{\beta\,\gamma}{2}+n\gamma \right )sin(t)\right )
$$
$$
\frac{1}{2}\left(\alpha^2+\beta^2-\frac {1}{2}
+\left (\beta^2-\alpha^2\right )sin(t)\right)\frac{1}{cos^2(t)}
+\frac{\gamma^2}{16}cos^2(t),
$$

$$
A(x)=x^3, \quad x(t)=\frac{4}{t^2}
$$\vspace{-5mm}
$$
V_n(t)=\frac{\gamma}{2}\left(\alpha+1 \right )
+\left(\frac {15}{4}+\alpha^2+4\,n\alpha+4\,\alpha+4\,n^2+8\,n \right)
\frac{1}{t^2}
+\frac{1}{4}\left (\alpha\;\beta+\frac{\gamma^2}{4} \right )\;t^2
$$
$$
+\frac{\beta\,\gamma}{16}{t}^{4}
+\frac{\beta^2}{64}t^6,
$$

$$     
A(x)=x^2(1-x), \quad x(t)=1-tanh^2(\frac{t}{2})
$$\vspace{-5mm}
$$
V_n(t)=\left(\frac{1}{cosh^2(t)-1}\right)
\left(-\left (2\,{n}^{2}+2+2\,n\alpha+\frac{\alpha^2}{2}+4\,n+\alpha\,\beta
+2\,\alpha+\frac{\alpha\,\gamma}{4}+2\,n\beta+2\,\beta\right )cosh(t)\right.
$$
$$
+\frac{1}{2}\left (\frac{-\gamma^2}{4}-\frac{\alpha\,\gamma}{2}
+{\frac {1}{2}}-\,\gamma+\,\alpha-\,\beta\,\gamma+\frac{\alpha^2}{2}\right )cosh^{2}
+\frac{\alpha\,\gamma}{4}cosh^{3}+\frac{\gamma^2}{16}cosh^4
$$
$$
+\left.\left(\frac{\gamma}{2}+4\,n+\alpha\,\beta
+\frac{3\;\alpha}{2}+{\beta}^{2}+2\,\beta+{\frac {3}{2}}+\frac{\alpha^2}{4}
+\frac{\alpha\,\gamma}{4}+\frac{\beta\,\gamma}{2}+\frac{\gamma^2}{16}
+2\,n\beta+2\,{n}^{2}+2\,n \alpha \right ) \right ),
$$

$$     
A(x)=x^4, \quad x(t)=\frac{1}{t}
$$\vspace{-5mm}
$$
V_n(t)=\left(\frac{\delta^2}{4}+\gamma+2\,n\gamma \right)
+\left (\gamma\,\delta+3\,n\beta+3\,\beta \right )t
+\left (\frac{3\beta\,\delta}{2}+{\gamma}^{2}\right ){t}^{2}
+3\,\beta\,\gamma\,{t}^{3}
+\frac{9\beta^2}{4}{t}^{4},
$$

$$     
A(x)=x^3(1-x), \quad x(t)=\frac{4}{4+t^2}
$$\vspace{-5mm}
$$
V_n(t)=-\left(\frac{\gamma}{2}+\delta+\beta\,\delta+\frac{\beta\,\gamma}{2}
+n\;\gamma \right )
+\left(\beta^2-\frac {1}{4}\right)\frac{1}{t^2}
$$
$$
+\frac{1}{2}\left (-n\delta+\frac{\delta^2}{2}-\frac{\beta\,\delta}{2}
-\delta+\frac{\gamma^2}{8}+\frac{\gamma\,\delta}{2} \right ){t}^{2}
+\frac{\delta}{8} \left (\frac{\gamma}{2}+\delta \right ){t}^{4}
+\frac {\delta^2}{64}\,t^6,
$$

$$     
A(x)=x^2(1+x^2), \quad x(t)=\frac{-1}{sinh(t)}
$$\vspace{-5mm}
$$
V_n(t)=\left (n+{n}^{2}-\frac{\gamma\,\delta}{2}+{\frac {1}{4}}+2\,n\beta+\beta
+{\beta}^{2}-\left (n\gamma+\gamma+\beta\,\gamma\right )sinh(t)\right )
$$
$$
+\left(\frac{\delta^2}{4}+\beta\,\delta\,sinh(t)-{\beta}^{2}
+\frac{1}{4} \right )\frac{1}{cosh^2(t)}
+\frac{\gamma^2}{4}cosh^2(t),
$$

$$     
A(x)=x^2(1-x^2), \quad x(t)=\frac{1}{cosh(t)}
$$\vspace{-4mm}
$$
V_n(t)=\left(\frac{1}{cosh^2(t)-1}\right)
\left(\left(\frac{-\gamma}{2}-n-\frac{\beta}{2}
+\frac{\gamma^2}{4}+\frac{\delta^2}{4}-\frac{\beta\,\gamma}{2}
-\frac{\beta\,\delta}{2}+\frac{\gamma\,\delta}{2}-\frac {1}{2}-n\beta
+\frac{\beta^2}{4}-n\gamma-n^2 \right) \right.
$$
$$
+\left (\frac{\beta\,\delta}{2}+n\beta-\frac{\gamma\,\delta}{2}
+n\gamma+\frac{\gamma}{2}+\frac {1}{4}+\frac{\beta}{2}+\frac{\beta\,\gamma}{2}
-\frac{\delta^2}{2}+{n}^{2}+\frac{\beta^2}{4}+n+\frac{\gamma^2}{4}
\right ){cosh(t)}^{2} 
$$
$$
+\left. \left (\frac{-\gamma\,\delta}{2}-n\delta-\delta
-\frac{\gamma^2}{2}-\frac{\beta\,\delta}{2}+\frac{\beta^2}{2} \right )cosh(t)
+\left ( \delta+n\;\delta+\frac{\beta\,\delta}{2}+\frac{\gamma\,\delta}{2}
\right )cosh^3(t)
+\frac{\delta^2}{4}cosh^4(t) \right),
$$

$$     
A(x)=x^2(1-x)^2, \quad x(t)=\frac{e^t}{1+e^t}
$$\vspace{-4mm}
$$
V_n(t)=\left(\frac{1}{\left(e^{-t}+1\right)^4\;S_6}\right)
\left(\frac{\delta^2}{4}\;S_4
+\delta \left(\delta-\frac{\beta}{2} \right )S_{5}\right.
$$
$$
+\left (\frac{1}{4}-\frac{\gamma\,\delta}{2}+n^2+\frac{\beta^2}{4}
-n\;\gamma+\frac{\beta}{2}+n-2\,\beta\,\delta+\frac{3\delta^2}{2}
+n\beta \right )S_{6}
$$
$$
\left (-3\,\beta\,\delta+\frac{\beta\,\gamma}{2}-3\,n\gamma+\gamma+4\,n
+{\delta}^{2}-2\,\gamma\,\delta+{\beta}^{2}+1+4\,{n}^{2}+2\,\beta+4\,n
\beta \right )S_{{7}}
$$
$$
+\left (3\,\beta+6\,n+\frac{3\beta^2}{2}-3\,\gamma\,
\delta+{\frac {3}{2}}+\frac{\delta^2}{4}+\frac{\gamma^2}{4}+6\,{n}^{2}-2
\,\beta\,\delta+4\,\gamma+6\,n\beta+2\,\beta\,\gamma-2\,n\gamma\right)S_{{8}}
$$
$$
+\left (4\,{n}^{2}-\frac{\beta\,\delta}{2}-2\,\gamma\,\delta+3\,\beta
\,\gamma+1+2\,n\gamma+4\,n+{\beta}^{2}+2\,\beta+{\gamma}^{2}+6\,\gamma
+4\,n\beta \right )S_{{9}}
$$
$$
+\left (\frac{-\gamma\,\delta}{2}+\frac{3\gamma^2}{2}
+4\,\gamma+n+2\,\beta\,\gamma+3\,n\gamma+{\frac{1}{4}}+\frac{\beta^2}{4}
+\frac{\beta}{2}+n\beta+{n}^{2}\right )S_{{10}}
$$
$$
+\left (n\gamma+{\gamma}^{2}+\frac{\beta\,\gamma}{2}+\gamma \right )S_{{11}}
\left.+\frac{\gamma^2}{4}S_{{12}} \right),
$$
where $S_k$ is defined as:
$$
S_k=exp \left(-2\;\left( \frac{k\;t}{2}+\left ( \gamma\;e^{-t}
+ \gamma+\delta \left ( e^t+1 \right) \right) \right) \right), \quad k=4,5,\cdots,12.
$$
\section{CONCLUSION}

As we saw by introducing of master function $A(x)$ as a polynomial of
order at most four, we could obtain all quasi-exactly second order differential
equations. It is shown that the eigen-equation relation $L\Psi(x)=E\Psi(x)$
generates a set of polynomials ${P_m(E)}$,
where these polynomials satisfy 3-term
and 4-term recursion relations for master function of at most three and four,
respectively. Finally the quasi-exactly solvablity leads to factorization of
polynomials $P_{n+N+1}(E)$ for $N\geq 0$ in terms of $P_{n+1}(E)$, where
by determining the roots of $P_{n+1}(E)$ we can determine first $n+1$
eigen-values of these quasi-exactly solvable differential equations.

\vspace{10mm}
\vspace{5mm}
{\bf {\Large  APPENDIX I}}

{\bf {\large  The First Five Polynomials $P_n(E)$, For $k=3$ }} 

To abbreviate, we set $F^{(i)}=(\frac{AW^\prime}{W})^{(i)}$,
 
$$
P_1(E)=-{\frac {E}{{\it F^{(0)}}}},
$$\vspace{15mm}
$$
P_2(E)=\frac{1}{2}{\frac {E{\it F^{(1)}}+{E}^{2}+{\it B^{(1)}}\,{\it F^{(0)}}}{{\it F^{(0)}}\,\left ({
\it A^{(1)}}+{\it F^{(0)}}\right )}},
$$\vspace{15mm}
$$
P_3(E)=\left(-{\it A^{(2)}}\,E{\it F^{(1)}}-{\it A^{(2)}}\,{E}^{2}-{\it A^{(2)}}\,{\it B^{(1)}}
\,{\it F^{(0)}}-2\,E{{\it F^{(1)}}}^{2}-3\,{\it F^{(1)}}\,{E}^{2}-2\,{\it F^{(1)}}\,{\it 
B^{(1)}}\,{\it F^{(0)}}-{E}^{3}\right.
$$\vspace{-9mm}
$$
\left. -3\,E{\it B^{(1)}}\,{\it F^{(0)}}+E{\it F^{(2)}}\,{\it A^{(1)}}+E{
\it F^{(2)}}\,{\it F^{(0)}}-2\,E{\it B^{(1)}}\,{\it A^{(1)}} \right)
$$\vspace{-9mm}
$$
\left / \left (6 F^{(0)}\,\left ( A^{(1)}
+ F^{(0)}\right )\left (2\, A^{(1)}+ F^{(0)}\right )\right)\right.,
$$\vspace{15mm}
$$
P_4(E)=-\left(-3\, {A^{(2)}}^2\;E^2-3\,{B^{(1)}}^2\;{F^{(0)}}^2
-6\,{B^{(1)}}^2\;F^{(0)}\,A^{(1)}+3\,F^{(2)}\,B^{(1)}\;
{F^{(0)}}^2+A^{(3)}\,B^{(1)}\,{F^{(0)}}^2\;\right.
$$\vspace{-9mm}
$$
+{\it A^{(3)}}\,{E}^{2}{\it F^{(0)}}
+2\,{\it A^{(3)}}\,{E}^{2}{\it A^{(1)}}-8\,{E}^{2}{\it B^{(1)}}\,{\it A^{(1)}}+4\,{E}^{2}{
\it F^{(2)}}\,{\it F^{(0)}}+7\,{E}^{2}{\it F^{(2)}}\,{\it A^{(1)}}-6\,{E}^{2}{\it B^{(1)}}\,{
\it F^{(0)}}
$$\vspace{-9mm}
$$
-6\,{{\it F^{(1)}}}^{2}{\it B^{(1)}}\,{\it F^{(0)}}-13\,{\it A^{(2)}}\,{\it F^{(1)}}\,{E
}^{2}
-9\,{\it A^{(2)}}\,E{{\it F^{(1)}}}^{2}-3\,{{\it A^{(2)}}}^{2}{\it B^{(1)}}\,{\it F^{(0)}}
-3\,{{\it A^{(2)}}}^{2}E{\it F^{(1)}}
$$\vspace{-9mm}
$$
-4\,{\it A^{(2)}}\,{E}^{3}-6\,E{{\it F^{(1)}}}^{3}-11
\,{{\it F^{(1)}}}^{2}{E}^{2}-6\,{\it F^{(1)}}\,{E}^{3}
-9\,{\it A^{(2)}}\,{\it F^{(1)}}\,{\it B^{(1)}}\,{\it F^{(0)}}
-6\,{\it A^{(2)}}\,E{\it B^{(1)}}\,{\it A^{(1)}}
$$\vspace{-9mm}
$$
+3\,{\it A^{(2)}}\,E{\it 
F^{(2)}}\,{\it F^{(0)}}+3\,{\it A^{(2)}}\,E{\it F^{(2)}}\,{\it A^{(1)}}-10\,{\it A^{(2)}}\,E{\it B^{(1)}}
\,{\it F^{(0)}}
-12\,{\it F^{(1)}}\,E{\it B^{(1)}}\,{\it A^{(1)}}+6\,{\it F^{(1)}}\,E{\it F^{(2)}}\,{
\it F^{(0)}}
$$\vspace{-9mm}
$$
+9\,{\it F^{(1)}}\,E{\it F^{(2)}}\,{\it A^{(1)}}-14\,{\it F^{(1)}}\,E{\it B^{(1)}}\,{
\it F^{(0)}}+2\,{\it A^{(3)}}\,{\it B^{(1)}}\,{\it F^{(0)}}\,{\it A^{(1)}}+{\it A^{(3)}}\,E{\it F^{(1)}}
\,{\it F^{(0)}}
$$\vspace{-9mm}
$$
\left.+2\,{\it A^{(3)}}\,E{\it F^{(1)}}\,{\it A^{(1)}}+6\,{\it F^{(2)}}
\,{\it B^{(1)}}\,{\it F^{(0)}}\,{\it A^{(1)}}-{E}^{4} \right)
$$\vspace{-9mm}
$$
\left/\left(24{\it F^{(0)}}\,\left ({\it A^{(1)}}+{\it F^{(0)}}\right )
\left (2\,{\it A^{(1)}}+{\it F^{(0)}}\right )\left (3\,{\it A^{(1)}}
+{\it F^{(0)}}\right )\right)\right.,
$$\vspace{15mm}
$$
P_5(E)=-\left(-46\,E{\it A^{(3)}}\,{\it B^{(1)}}\,{\it F^{(0)}}\,{\it A^{(1)}}
-88\,E{\it F^{(2)}}\,{\it B^{(1)}}\,{\it F^{(0)}}\,{\it A^{(1)}}
+16\,{\it A^{(3)}}\,E{\it F^{(2)}}
\,{\it A^{(1)}}\,{\it F^{(0)}}\right.
$$\vspace{-9mm}
$$
+24\,E{{\it F^{(2)}}}^{2}{\it A^{(1)}}\,{\it F^{(0)}}
-48\,{\it F^{(2)}}\,E{\it B^{(1)}}\,{{\it A^{(1)}}}^{2}+18\,E{{\it F^{(2)}}}^{2}{{\it A^{(1)}}}^{2}
+6\,E{{\it F^{(2)}}}^{2}{{\it F^{(0)}}}^{2}+24\,E{{\it B^{(1)}}}^{2}{{\it A^{(1)}}}^{2}
$$\vspace{-9mm}
$$
+12\,{\it 
A^{(3)}}\,E{\it F^{(2)}}\,{{\it A^{(1)}}}^{2}+4\,{\it A^{(3)}}\,E{\it F^{(2)}}\,{{\it F^{(0)}}}^{2}
-24\,{\it A^{(3)}}\,E{\it B^{(1)}}\,{{\it A^{(1)}}}^{2}-32\,{\it F^{(1)}}\,{\it A^{(3)}}\,{\it 
B^{(1)}}\,{\it F^{(0)}}\,{\it A^{(1)}}
$$\vspace{-9mm}
$$
-60\,{\it F^{(1)}}\,{\it F^{(2)}}\,{\it B^{(1)}}\,{\it F^{(0)}}\,{
\it A^{(1)}}+50\,E{{\it B^{(1)}}}^{2}{\it F^{(0)}}\,{\it A^{(1)}}-25\,E{\it F^{(2)}}\,{\it B^{(1)}}
\,{{\it F^{(0)}}}^{2}
-13\,E{\it A^{(3)}}\,{\it B^{(1)}}\,{{\it F^{(0)}}}^{2}
$$\vspace{-9mm}
$$
-46\,{\it F^{(1)}}
\,{\it A^{(3)}}\,{E}^{2}{\it A^{(1)}}+80\,{\it F^{(1)}}\,{E}^{2}{\it B^{(1)}}\,{\it A^{(1)}}-40
\,{\it F^{(1)}}\,{E}^{2}{\it F^{(2)}}\,{\it F^{(0)}}-91\,{\it F^{(1)}}\,{E}^{2}{\it F^{(2)}}\,{
\it A^{(1)}}
$$\vspace{-9mm}
$$
+50\,{\it F^{(1)}}\,{E}^{2}{\it B^{(1)}}\,{\it F^{(0)}}-54\,{\it A^{(2)}}\,{\it F^{(1)}}
\,E{\it F^{(2)}}\,{\it F^{(0)}}-84\,{\it A^{(2)}}\,{\it F^{(1)}}\,E{\it F^{(2)}}\,{\it A^{(1)}}+137
\,{\it A^{(2)}}\,{\it F^{(1)}}\,E{\it B^{(1)}}\,{\it F^{(0)}}
$$\vspace{-9mm}
$$
-24\,{\it A^{(2)}}\,{\it A^{(3)}}\,{
\it B^{(1)}}\,{\it F^{(0)}}\,{\it A^{(1)}}
-10\,{\it A^{(2)}}\,{\it A^{(3)}}\,E{\it F^{(1)}}\,{\it F^{(0)}
}-24\,{\it A^{(2)}}\,{\it A^{(3)}}\,E{\it F^{(1)}}\,{\it A^{(1)}}
$$\vspace{-9mm}
$$
-54\,{\it A^{(2)}}\,{\it F^{(2)}}\,
{\it B^{(1)}}\,{\it F^{(0)}}\,{\it A^{(1)}}+20\,{\it F^{(1)}}\,{{\it B^{(1)}}}^{2}{{\it F^{(0)}}}^{2
}-5\,{\it A^{(3)}}\,{E}^{3}{\it F^{(0)}}-14\,{\it A^{(3)}}\,{E}^{3}{\it A^{(1)}}+20\,{E}^{
3}{\it B^{(1)}}\,{\it A^{(1)}}
$$\vspace{-9mm}
$$
-10\,{E}^{3}{\it F^{(2)}}\,{\it F^{(0)}}-25\,{E}^{3}{\it F^{(2)}}
\,{\it A^{(1)}}+10\,{E}^{3}{\it B^{(1)}}\,{\it F^{(0)}}+15\,E{{\it B^{(1)}}}^{2}{{\it F^{(0)}}}
^{2}
+66\,{\it A^{(2)}}\,{E}^{2}{\it B^{(1)}}\,{\it A^{(1)}}
$$\vspace{-9mm}
$$
-33\,{\it A^{(2)}}\,{E}^{2}{
\it F^{(2)}}\,{\it F^{(0)}}-63\,{\it A^{(2)}}\,{E}^{2}{\it F^{(2)}}\,{\it A^{(1)}}+50\,{\it A^{(2)}}
\,{E}^{2}{\it B^{(1)}}\,{\it F^{(0)}}+72\,{\it A^{(2)}}\,{{\it F^{(1)}}}^{2}{\it B^{(1)}}\,{
\it F^{(0)}}
$$\vspace{-9mm}
$$
+72\,{{\it F^{(1)}}}^{2}E{\it B^{(1)}}\,{\it A^{(1)}}-36\,{{\it F^{(1)}}}^{2}E{\it 
F^{(2)}}\,{\it F^{(0)}}-72\,{{\it F^{(1)}}}^{2}E{\it F^{(2)}}\,{\it A^{(1)}}+70\,{{\it F^{(1)}}}^{2}
E{\it B^{(1)}}\,{\it F^{(0)}}
$$\vspace{-9mm}
$$
-12\,{\it A^{(3)}}\,E{{\it F^{(1)}}}^{2}{\it F^{(0)}}
-32\,{\it A^{(3)}}\,E{{\it F^{(1)}}}^{2}{\it A^{(1)}}
+48\,{\it F^{(1)}}\,{{\it B^{(1)}}}^{2}{\it F^{(0)}}\,{\it 
A^{(1)}}-24\,{\it F^{(1)}}\,{\it F^{(2)}}\,{\it B^{(1)}}\,{{\it F^{(0)}}}^{2}
$$\vspace{-9mm}
$$
-12\,{\it F^{(1)}}\,{
\it A^{(3)}}\,{\it B^{(1)}}\,{{\it F^{(0)}}}^{2}
-17\,{\it F^{(1)}}\,{\it A^{(3)}}\,{E}^{2}{\it F^{(0)}}
-24\,{\it A^{(2)}}\,{\it F^{(2)}}\,{\it B^{(1)}}\,{{\it F^{(0)}}}^{2}-10\,{\it A^{(2)}}\,{
\it A^{(3)}}\,{E}^{2}{\it F^{(0)}}
$$\vspace{-9mm}
$$
-24\,{\it A^{(2)}}\,{\it A^{(3)}}\,{E}^{2}{\it A^{(1)}}-10\,{
\it A^{(2)}}\,{\it A^{(3)}}\,{\it B^{(1)}}\,{{\it F^{(0)}}}^{2}
+108\,{\it A^{(2)}}\,{\it F^{(1)}}\,E{\it B^{(1)}}\,{\it A^{(1)}}
$$\vspace{-9mm}
$$
+10\,{\it F^{(1)}}\,{E}^{4}+18\,{{\it A^{(2)}}}^{3}{E}^{2}+27
\,{{\it A^{(2)}}}^{2}{E}^{3}+10\,{\it A^{(2)}}\,{E}^{4}+24\,E{{\it F^{(1)}}}^{4}
+50\,{{\it F^{(1)}}}^{3}{E}^{2}
$$\vspace{-9mm}
$$
+35\,{{\it F^{(1)}}}^{2}{E}^{3}+{E}^{5}+93\,{{\it A^{(2)}}}
^{2}{\it F^{(1)}}\,{E}^{2}+66\,{{\it A^{(2)}}}^{2}E{{\it F^{(1)}}}^{2}+18\,{{\it A^{(2)}}}
^{3}{\it B^{(1)}}\,{\it F^{(0)}}+18\,{{\it A^{(2)}}}^{3}E{\it F^{(1)}}
$$\vspace{-9mm}
$$
+22\,{\it A^{(2)}}\,{{\it B^{(1)}}}^{2}{{\it F^{(0)}}}^{2}
+72\,{\it A^{(2)}}\,E{{\it F^{(1)}}}^{3}+127\,{\it A^{(2)}}
\,{{\it F^{(1)}}}^{2}{E}^{2}+65\,{\it A^{(2)}}\,{\it F^{(1)}}\,{E}^{3}
$$\vspace{-9mm}
$$
+24\,{{\it F^{(1)}}}
^{3}{\it B^{(1)}}\,{\it F^{(0)}}
+66\,{{\it A^{(2)}}}^{2}{\it F^{(1)}}\,{\it B^{(1)}}\,{\it F^{(0)}}+36\,{{\it A^{(2)}}}^{2}E{\it B^{(1)}}
\,{\it A^{(1)}}
$$\vspace{-9mm}
$$
\left.-18\,{{\it A^{(2)}}}^{2}E{\it F^{(2)}}\,{
\it F^{(0)}}-18\,{{\it A^{(2)}}}^{2}E{\it F^{(2)}}\,{\it A^{(1)}}+63\,{{\it A^{(2)}}}^{2}E{\it 
B^{(1)}}\,{\it F^{(0)}}+48\,{\it A^{(2)}}\,{{\it B^{(1)}}}^{2}{\it F^{(0)}}\,{\it A^{(1)}}\right)
$$\vspace{-9mm}
$$
\left/\left(120{\it F^{(0)}}
\,\left ({\it A^{(1)}}+{\it F^{(0)}}\right )\left (2\,{\it A^{(1)}}+{\it F^{(0)}}\right )
\left (3\,{\it A^{(1)}}+{\it F^{(0)}}\right )\left (4\,{\it A^{(1)}}+{\it F^{(0)}}\right)
\right)\right..
$$

\vspace{10mm}
{\bf {\Large  APPENDIX II}}

{\bf {\large  The First Four Polynomials $P_n(E)$, For $k=4$ }}

$$                                                
P_{1}(E)=\frac{E}{F^{(0)}},
$$\vspace{15mm}
$$
P_2(E)=-\frac{1}{2}\frac{E{\it F^{(1)}}+{E}^{2}-{\it B^{(1)}}\,{\it F^{(0)}}}
{{\it F^{(0)}}\left ({ \it A^{(1)}}+{\it F^{(0)}}\right )},
$$\vspace{15mm}
$$
P_3(E)=\left({\it A^{(2)}}\,E{\it F^{(1)}}+{\it A^{(2)}}\,{E}^{2}-{\it A^{(2)}}\,{\it B^{(1)}}
\,{\it F^{(0)}}+2\,E{{\it F^{(1)}}}^{2}+3\,{\it F^{(1)}}\,{E}^{2}-2\,{\it F^{(1)}}\,{\it 
B^{(1)}}\,{\it F^{(0)}}\right.
$$\vspace{-9mm}
$$
+\left.{E}^{3}+E{\it B^{(1)}}\,{\it F^{(0)}}
-E{\it F^{(2)}}\,{\it A^{(1)}}-E{\it F^{(2)}
}\,{\it F^{(0)}}+2\,E{\it B^{(1)}}\,{\it A^{(1)}}+{\it B^{(2)}}\,{\it F^{(0)}}\,{\it A^{(1)}}+{\it 
B^{(2)}}\,{{\it F^{(0)}}}^{2}\right)
$$\vspace{-9mm}
$$
\left/\left(6{\it F^{(0)}}\,\left ({\it A^{(1)}}+{\it F^{(0)}}\right )
\left (2\,{\it A^{(1)}}+{\it F^{(0)}}\right )\right)\right.,
$$\vspace{15mm}
$$
P_4(E)=-\left(3\,{\it A^{(2)}}\,{\it B^{(2)}}\,{{\it F^{(0)}}}^{2}+{\it A^{(3)}}\,{\it B^{(1)}
}\,{{\it F^{(0)}}}^{2}+9\,{\it A^{(2)}}\,E{{\it F^{(1)}}}^{2}-3\,{\it A^{(2)}}\,E{\it F^{(2)}}
\,{\it A^{(1)}}\right.
$$\vspace{-9mm}
$$
+2\,{\it A^{(2)}}\,E{\it B^{(1)}}\,{\it F^{(0)}}-9\,{\it A^{(2)}}\,{\it F^{(1)}}\,{
\it B^{(1)}}\,{\it F^{(0)}}-2\,E{\it B^{(2)}}\,{{\it F^{(0)}}}^{2}+3\,{\it A^{(2)}}\,{\it B^{(2)}}\,
{\it F^{(0)}}\,{\it A^{(1)}}+6\,{\it A^{(2)}}\,E{\it B^{(1)}}\,{\it A^{(1)}}
$$\vspace{-9mm}
$$
-3\,{\it A^{(2)}}\,E{
\it F^{(2)}}\,{\it F^{(0)}}-9\,{\it F^{(1)}}\,E{\it F^{(2)}}\,{\it A^{(1)}}+4\,{\it F^{(1)}}\,E{\it 
B^{(1)}}\,{\it F^{(0)}}-3\,{{\it B^{(1)}}}^{2}{{\it F^{(0)}}}^{2}-8\,E{\it B^{(2)}}\,{\it F^{(0)}}\,
{\it A^{(1)}}
$$\vspace{-9mm}
$$
-2\,{\it A^{(3)}}\,E{\it F^{(1)}}\,{\it A^{(1)}}-{\it A^{(3)}}\,E{\it F^{(1)}}\,{\it F^{(0)}
}+2\,{\it A^{(3)}}\,{\it B^{(1)}}\,{\it F^{(0)}}\,{\it A^{(1)}}+6\,{\it F^{(2)}}\,{\it B^{(1)}}\,{
\it F^{(0)}}\,{\it A^{(1)}}+3\,E{\it F^{(3)}}\,{\it A^{(1)}}\,{\it F^{(0)}}
$$\vspace{-9mm}
$$
-6\,{\it F^{(1)}}\,E{\it 
F^{(2)}}\,{\it F^{(0)}}+3\,{{\it A^{(2)}}}^{2}{E}^{2}+4\,{\it A^{(2)}}\,{E}^{3}+6\,E{{\it 
F^{(1)}}}^{3}+11\,{{\it F^{(1)}}}^{2}{E}^{2}+6\,{\it F^{(1)}}\,{E}^{3}+3\,{{\it A^{(2)}}}^
{2}E{\it F^{(1)}}
$$\vspace{-9mm}
$$
-3\,{{\it A^{(2)}}}^{2}{\it B^{(1)}}\,{\it F^{(0)}}+13\,{\it A^{(2)}}\,{\it F^{(1)}
}\,{E}^{2}-6\,{{\it F^{(1)}}}^{2}{\it B^{(1)}}\,{\it F^{(0)}}+3\,{\it F^{(1)}}\,{\it B^{(2)}}\,
{{\it F^{(0)}}}^{2}+4\,{E}^{2}{\it B^{(1)}}\,{\it F^{(0)}}
$$\vspace{-9mm}
$$
-7\,{E}^{2}{\it F^{(2)}}\,{\it 
A^{(1)}}-4\,{E}^{2}{\it F^{(2)}}\,{\it F^{(0)}}+8\,{E}^{2}{\it B^{(1)}}\,{\it A^{(1)}}-2\,{\it 
A^{(3)}}\,{E}^{2}{\it A^{(1)}}-{\it A^{(3)}}\,{E}^{2}{\it F^{(0)}}+3\,{\it F^{(2)}}\,{\it B^{(1)}}\,
{{\it F^{(0)}}}^{2}
$$\vspace{-9mm}
$$
-6\,{{\it B^{(1)}}}^{2}{\it F^{(0)}}\,{\it A^{(1)}}+2\,E{\it F^{(3)}}\,{{
\it A^{(1)}}}^{2}+E{\it F^{(3)}}\,{{\it F^{(0)}}}^{2}-6\,E{\it B^{(2)}}\,{{\it A^{(1)}}}^{2}+{E
}^{4}+3\,{\it F^{(1)}}\,{\it B^{(2)}}\,{\it F^{(0)}}\,{\it A^{(1)}}
$$\vspace{-9mm}
$$
\left.+12\,{\it F^{(1)}}\,E{\it B^{(1)}}\,{\it A^{(1)}}\right)
\left/\left(24{\it F^{(0)}}\,\left ({\it A^{(1)}}+{\it F^{(0)}}\right )
\left (2\,{\it A^{(1)}}+{\it F^{(0)}}\right )
\left (3\,{\it A^{(1)}}+{\it F^{(0)}}\right )
\right)\right.
$$

\vspace{10mm}
{\bf {\large  ACKNOWLEDGEMENT}}

We wish to thank  Dr. S. K. A. Seyed Yagoobi for his careful reading the 
article and for his constructive comments.

\end{document}